\documentclass[twocolumn,aps,prl]{revtex4}
\usepackage{mathrsfs}
\usepackage{amsmath}
\usepackage{amsfonts}
\usepackage{amssymb}
\usepackage{amsthm}
\usepackage{graphicx}

\begin{document}

\title{Domain wall propagation due to the synchronization with circularly
polarized microwaves}
\author{P. Yan}
\affiliation{Physics Department, The Hong Kong University of Science and Technology,
Clear Water Bay, Hong Kong SAR, China}
\author{X. R. Wang}
\affiliation{Physics Department, The Hong Kong University of Science and Technology,
Clear Water Bay, Hong Kong SAR, China}

\begin{abstract}
Finding a new control parameter for magnetic domain wall (DW) motion in
magnetic nanostructures is important in general and in particular for the
spintronics applications. Here, we show that a circularly polarized magnetic
field (CPMF) at GHz frequency (microwave) can efficiently drive a DW to
propagate along a magnetic nanowire. Two motion modes are identified:
rigid-DW propagation at low frequency and oscillatory propagation at high
frequency. Moreover, DW motion under a CPMF is equivalent to the DW motion
under a uniform spin current in the current perpendicular to the plane
magnetic configuration proposed recently by Khvalkovskiy \emph{et al.}
[Phys. Rev. Lett. \textbf{102}, 067206 (2009)], and the CPMF frequency plays
the role of the current.
\end{abstract}

\maketitle





Controlled manipulation of magnetic domain-wall (DW) propagation in
nanowires has spurred intensive research in recent years\cite%
{Walker,Cowburn,Erskine,xrw,Slon,
szhang,szhang1,Fert,Parkin1,Parkin2,Li,Khvalkovskiy} in nano-magnetism
because of its fundamental interest and the potential impact in spintronic
device technology. Both static magnetic fields\cite%
{Walker,Cowburn,Erskine,xrw} and electric currents\cite%
{Slon,szhang,szhang1,Fert,Parkin1, Parkin2,Li,Khvalkovskiy} can be the
control parameters. However, their control mechanisms are different, and
they have different advantages and disadvantages. Thus, finding any new
control parameter for DW motion should be of great interest.

Although it is known both theoretically\cite{ZZSun} and experimentally\cite%
{Xiao} that a microwave can be used to manipulate the motion of single
domain magnetic particles (macrospins or Stoner particles), a microwave was
hardly thought to be an effective control parameter for DW motion. In this
letter, we show that a circularly polarized magnetic field (CPMF) at GHz
frequency (microwave) can efficiently drive a magnetic DW to propagate along
a nanowire at a high speed. Similar to a spin-polarized current, a CPMF can
generate both Slonczewski-like and field-like spin-transfer torques (STTs)
inside a DW. Unlike the STTs generated by a spin-polarized current, the
field-like STT is much bigger than the Slonczewski-like STT. Thus, a CPMF
generates more useful STT\cite{szhang1} in comparison with that of a
spin-polarized current. Moreover, its driving mechanism is very different
from those of a static axial magnetic field and a spin-polarized current.

In order to appreciate the CPMF-driven DW motion, let us recall the driving
mechanisms of a static axial magnetic field and a spin-polarized current. A
DW propagates along a wire under a static magnetic field because the Zeeman
energy must be released to compensate the dissipated energy due to the
non-existence\cite{note1} of a static DW in a static magnetic field. An
electric current, on the other hand, moves DWs by the STT due to the
transfer of spin angular momentum from conduction electrons to local spins%
\cite{Slon}. A spin-polarized current can transfer two types of torques to a
local magnetization $\vec{M}=M_s\vec{m}$. One is the Slonczewski-like STT
\cite{Slon} of $b\vec{m}\times (\vec{m}\times\vec{s})$ (b-term), where $\vec{%
s}$ is the polarization direction of the spin-polarized current. The other
one is a field-like torque\cite{szhang} of $c \vec{m}\times\vec{s}$
(c-term). $b$ and $c$ are parameters roughly proportional to current\cite%
{Slon,szhang}. The effects of b- and c-terms on DW propagation are very
different. b-term is incapable of generating a sustained wall motion, except
at very large current, while c-term can drive a DW to propagate along the
carrier direction\cite{szhang1}. Unfortunately, c-term is normally much
smaller than b-term\cite{Li}. A large current density is needed to achieve a
technologically useful DW propagation velocity\cite{Parkin1}, but the
associated Joule heating could affect device performance. Thus, it should be
interesting if one can generate a large c-term (in comparison with b-term)
either in a new architecture or by using a new control parameter. One
solution along the first line of the thinking is provided by Khvalkovskiy
\textit{et al.}\cite{Khvalkovskiy} who proposed a current perpendicular to
the plane magnetic configuration in a sandwiched magnetic nanowire
structure. Here we provide a solution along the second line of the thinking.

We consider a head-to-head (HH) DW in a magnetic nanowire whose easy axis is
along the wire axis defined as the z-axis shown in Fig. 1. The motion of the
magnetization $\vec{M}=M_s\vec{m}$, is governed by the
Landau-Lifshitz-Gilbert (LLG) equation\cite{gilbert}
\begin{equation}
\frac{\partial\vec{m}}{\partial t}=-\gamma\vec{m}\times\vec{h}_{eff} +\alpha%
\vec{m}\times\frac{\partial\vec{m}}{\partial t},  \label{LLG}
\end{equation}%
here $\vec{h}_{eff}=-\frac{1}{\mu_0}\delta U/\delta\vec{M}$ is the effective
magnetic field that is the variational derivative of the free energy density
$U[\vec{M}]$ with respect to magnetization $\vec{M}$, $\gamma=g\left\vert
e\right\vert /2m_e$ is the gyromagnetic ratio, $\mu_0$ is the vacuum
magnetic permeability and $\alpha$ is the phenomenological Gilbert damping
constant\cite{gilbert} which measures dissipative effect. Eq. \eqref{LLG} is
a nonlinear partial differential equation that can be solved exactly only in
some special cases\cite{Walker,xrw1}.

A macrospin can synchronize its motion with a CPMF. A synchronized motion
generates a damping field that forces the spin to move perpendicularly to
the synchronized motion, leading to a dramatic effect of a CPMF on a Stoner
particle\cite{ZZSun,Xiao}. Thus, it is natural to study the CPMF effect on
DW motion since a DW may also synchronize its motion with a CPMF due to the
DW texture nature if the motion of its constitute local spins are
synchronized. To demonstrate that this can indeed happen, let us study the
motion of a HH DW in a uniaxial wire under a CPMF $\vec{h}(t)=
h_0(\cos\omega t\hat{x}+\sin \omega t\hat{y})$ with frequency $\omega $ and
amplitude $h_0$. The free energy density $U(M_z)$ is a functional function
of $M_z$, and the effective field takes a form of $\vec{h}_ {eff}=\vec{h}%
(t)+f(M_z)\hat{z}+\frac{2A}{\mu_0M_s^2}\frac{\partial^2 \vec{M}}{\partial z^2%
},$ where $f$ is the anisotropy field and the last term is the exchange
field with the exchange coefficient $A$. The physics is that DW synchronize
its motion with a CPMF so that DW plane rotates around wire axis at angular
velocity $\omega$. According to the LLG equation (Eq. \eqref{LLG}), the
precession motion around $z-$axis gives rise to a damping torque $\mathbf{T}%
_d=\alpha\left(\vec{m}\times\frac{\partial\vec{m}}{\partial t}
\right)=-\alpha\omega\sin\theta \vec{e}_\theta$ ($\vec{e}_\theta$ is unit
vector of $\theta$ direction) shown in Fig. 1. Also, because of the lag of
the DW motion with the field ($\delta $ is the angle between the DW plane
and CPMF plane defined by $\vec{h}_{0}$ and $z-$ axis), the field exerts a
torque with $\theta$-component $T_p=-\gamma\left( \vec{m} \times \vec{h}%
_{eff}\right) \cdot \vec{e}_\theta=\gamma h_0\sin\delta$ shown in Fig. 1.
Later analysis (Eq. \eqref{DWmotion} derived later) shows that $T_p$ always
overwhelms $T_d$ so that the HH DW propagates to the left along the wire (a
collective motion of spins along $\theta$-direction corresponds to the DW
propagation along the wire).
\begin{figure}[tbph]
\begin{center}
\includegraphics[width=8.5cm, height=2.cm]{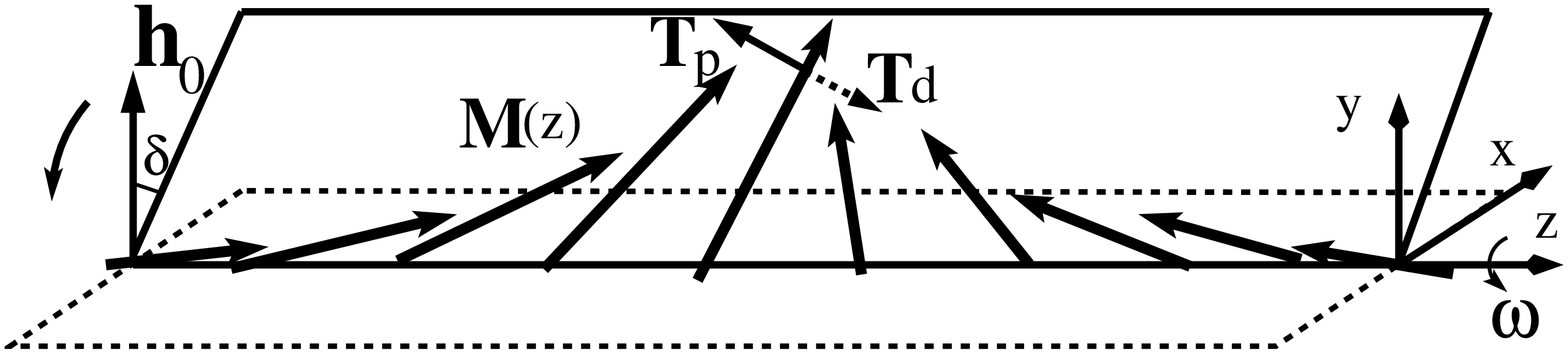}
\end{center}
\caption{Dashed arrow denotes the damping torque $T_{d}$ on the local spins
(long arrows) when the DW plane moves around (indicated by the curved arrow)
the wire axis during synchronization. Short arrow denotes torque $T_p$
arising from the lag of DW motion and the field.}
\label{fig1}
\end{figure}

For the motion of a HH DW under the CPMF, it is convenient to use the
rotation frame of
\begin{equation}
\left(
\begin{array}{c}
\hat{x}^{\prime } \\
\hat{y}^{\prime } \\
\hat{z}^{\prime }%
\end{array}%
\right) =\left(
\begin{array}{ccc}
\cos \omega t & \sin \omega t & 0 \\
-\sin \omega t & \cos \omega t & 0 \\
0 & 0 & 1%
\end{array}%
\right) \left(
\begin{array}{c}
\hat{x} \\
\hat{y} \\
\hat{z}%
\end{array}%
\right) ,
\end{equation}%
where $\hat{x}^{\prime },\ \hat{y}^{\prime },\ \hat{z}^{\prime }$ are the
unit vectors of Cartesian coordinates in the rotation frame. $\vec{m}$
becomes $\vec{m}^\prime=\left(
\begin{array}{ccc}
\sin \theta\cos\eta, & \sin\theta \sin\eta , & \cos\theta%
\end{array}
\right)$ in the rotation frame, where $\eta=-\delta$ is the azimuthal angle
of $\vec{m}$ in the rotation frame which is related to azimuthal angle $\phi
$ in the laboratory frame by $\eta =\phi -\omega t$. Polar angle $\theta$ is
the same in the two frames. The time derivatives of $\vec{m}$ in the two
reference frames are connected to each other by $\frac{\partial\vec{m}}{%
\partial t}=\frac{\partial \vec {m}^{\prime }}{\partial t}-\vec{m}^{\prime
}\times \vec{\omega}$, here angular velocity vector $\vec{\omega}=\omega
\hat{z}$ is assumed (Fig. 1). Eq. \eqref{LLG} in this rotation frame becomes
\begin{equation}
\frac{\partial \vec{m}^{\prime }}{\partial t}=-\gamma \vec{m}^{\prime
}\times \left( \vec{h}_{eff}^{\prime }+\frac{\alpha }{\gamma }\vec{m}
^\prime\times\vec{\omega}-\frac{\vec{\omega}}{\gamma }\right) +\alpha \vec{m}%
^\prime\times \frac{\partial \vec{m}^\prime }{\partial t},  \label{LLG1}
\end{equation}
where the effective field is $\vec{h}_{eff}^{\prime }=h_0\hat{x}^\prime
+f(\cos\theta)\hat{z}^\prime+\frac{2A}{\mu_0M_s}\frac{\partial^2 \vec{m}%
^\prime}{\partial z^{\prime 2}}$ that does not depend on time (In general,
the field due to the magnetic anisotropy depends on time in a rotation
frame, but it is time-independent for the uniaxial wire).

Eq. \eqref{LLG1} does not have any explicit time-dependent term, and the
original problem becomes a DW subjected to a transverse field $h_0\hat{x}
^\prime$, a Slonczewski-like torque (b-term) $\alpha\vec{m}^\prime\times (%
\vec{m}^\prime\times\vec{\omega})$, and a field-like torque (c-term) $\vec{m}%
^{\prime }\times \vec{\omega}.$ Thus a DW under a CPMF behaves like the DW
under a current-induced STT if one views $\vec{\omega}$ as a spin polarized
current. Unlike the STT from a real spin-polarized current, c-term is much
larger than b-term. Thus a CPMF is very efficient in driving DW propagation
along the wire. It should be noticed that the equivalent spin-polarized
current is uniformly applied to a DW, instead of spatially dependent
spin-polarized current related to $\frac{\partial \vec{m}}{\partial z}$
inside a DW\cite{szhang,szhang1}. Interestingly, Eq. \eqref{LLG1} is exactly
the same as that in a recently studied system\cite{Khvalkovskiy} of
current-driven DW motion in a sandwiched long and narrow spin valve, in
which the magnetization of reference magnetic layer plays the role of field
polarity $\vec{\omega}$. Thus a DW of uniaxial wire under a CPMF is
equivalent to the DW in composite spin valves under a current\cite%
{Khvalkovskiy}.

Early study\cite{Khvalkovskiy} on Eq. \eqref{LLG1} showed that DW propagates
like a rigid body under small torques, corresponding to a perfect
synchronized motion with the CPMF in the current case. It is noted that $%
\vec{\digamma}=\frac{\alpha }{\gamma }\vec{m}^{\prime }\times \vec{\omega}$
is a non-conservative field since $\nabla _{\vec{m}^{\prime }}\times \vec{%
\digamma}=-\frac{2\alpha }{\gamma }\vec{\omega}\neq 0.$
In order to find out how the DW propagation velocity depends on the
amplitude and the frequency of the CPMF, we adopt the generalized analysis
of Schryer and Walker\cite{Walker}. It is required to first find the static
DW solutions of Eq. \eqref{LLG1} at $\vec{\omega}=0$ (the case of a constant
transverse field $\vec{h}_{0}$). For the conventional uniaxial anisotropy $%
U(\cos \theta )=-\frac{1}{2}K\cos ^{2}\theta ,$ a static DW centered at $%
z^{\prime }=Q$ exists with following DW profile $\theta =\theta \left(
z^{\prime }-Q\right) ,\eta =0,$ where\cite{Vladimir}
\begin{equation}
\sin \theta \left( z^{\prime }\right) =\sin \theta _{0}+\frac{\cos
^{2}\theta _{0}}{\cosh \left[ \frac{z^{\prime }}{\Delta }\cos \theta _{0}%
\right] +\sin \theta _{0}},  \label{profile}
\end{equation}%
here $\theta_0=\sin^{-1}\left( \mu_0M_sh_0/K\right)$ is the tilted polar
angle of two domains due to transverse magnetic field $\vec{h}_0,$ and $%
\Delta=\sqrt{2A/K}$ is the static DW width without external magnetic field.
Assume the profile of the moving DW is the same as that of the static one,
then the moving DW is given by Eq. \eqref{profile} with collective
coordinates\cite{Vladimir,Tatara,Li1} $\left(Q,\text{ }\eta \right) $ being
functions of time. Substituting Eq. \eqref{profile} into Eq. \eqref{LLG1},
the equations for $Q$ and $\eta $ are
\begin{eqnarray}
\dot{\eta} &=&-\omega -\frac{\alpha \gamma h_{0}}{1+\alpha ^{2}}\sin \eta ,
\label{DWmotion} \\
\dot{Q} &=&-\frac{\omega +\dot{\eta}}{\alpha \left( 1-\sin \theta
_{0}\right) }\Delta .  \label{Oscillation}
\end{eqnarray}%
Solution of Eq. \eqref{DWmotion} with initial condition $\eta =0$ is
\begin{equation}
\cot \frac{\eta }{2}=%
\begin{cases}
-\frac{\omega _{c}}{\omega }-\sqrt{\left( \frac{\omega _{c}}{\omega }\right)
^{2}-1}\coth \frac{\sqrt{\omega _{c}^{2}-\omega ^{2}}}{2}t, & \omega <\omega
_{c} \\
-\frac{\omega _{c}}{\omega }-\sqrt{1-\left( \frac{\omega _{c}}{\omega }%
\right) ^{2}}\cot \frac{\sqrt{\omega ^{2}-\omega _{c}^{2}}}{2}t, & \omega
>\omega _{c}%
\end{cases}%
\end{equation}%
where $\omega_c=\frac{\alpha\gamma h_0}{1+\alpha^2}$ is the critical
frequency that separates two modes: $\eta $ approaches exponentially a fixed
value $\sin^{-1}\left(-\omega /\omega_c\right) $ in a time scale of $1/\sqrt{%
\omega _{c}^{2}-\omega ^{2}}$ for $\omega<\omega_c$. This is the fully
synchronized motion. Since $T_p=\left(\alpha+\frac{1} {\alpha}\right)\omega$
is always larger than $T_d=\alpha\omega\sin\theta$, the HH DW propagates to
the left as mentioned early. $\eta $ keeps rotating around z-axis with a
period of $2\pi/\sqrt{\omega^2-\omega_c^2}$ and a variable velocity for $%
\omega>\omega_c$, corresponding to an incomplete synchronization. For $%
\omega <\omega _{c}$, the steady DW propagation velocity is given by Eq. %
\eqref{Oscillation} when $\dot{\eta}=0$ is reached,
\begin{equation}
\upsilon=\dot{Q}\left( t\rightarrow\infty\right)=-\frac{\omega\Delta} {%
\alpha \left( 1-\sin \theta _{0}\right) }.  \label{velocity}
\end{equation}
Similar to the low current velocity, DW propagation velocity is linear in $%
\omega $ as shown by the solid lines in Fig. 2 for low frequency. For $%
\omega>\omega_c,$ the precession velocity of DW plane change with time.
According to Eq. \eqref{Oscillation}, DW velocity will also change with
time. Since the average angular velocity of $\eta$ is $\Omega=\sqrt{\omega
^{2}-\omega _{c}^{2}},$ the averaged DW velocity is
\begin{equation}
\bar{\upsilon}=-\frac{\omega -\sqrt{\omega ^{2}-\omega _{c}^{2}}}{\alpha
\left( 1-\sin \theta _{0}\right) }\Delta ,  \label{decay}
\end{equation}%
monotonically decreasing with $\omega $ shown by the solid curves in Fig. 2
for large $\omega $. The results can also be understood from the energy
consideration\cite{xrw}. For $\omega <\omega _{c}$, the non-conservative
field does not do any work to the system for a rigid DW propagation, and the
energy dissipation must be compensated by the energy released from the DW
propagation\cite{xrw}. According to Ref. \cite{xrw}, the DW velocity is
proportional to DW width and the axial field which is the frequency in the
present case. For $\omega>\omega_c$, DW plane precesses around the wire
while it propagates along the wire. The non-conservative field pumps energy
to the system so that the system needs to release less energy, corresponding
to a low DW velocity shown in Eq. \eqref{decay}.
\begin{figure}[tbph]
\begin{center}
\includegraphics[width=9cm, height=7cm]{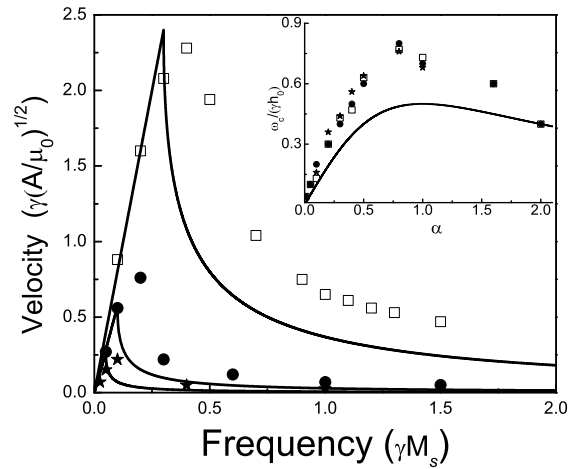}
\end{center}
\caption{DW velocity versus CPMF frequency with $\protect\alpha =0.1$ and $%
K=8$. Symbols are numerical data for field amplitude $h_{0}=3$ (squares), $1$
(circles), and $0.5$ (stars). Solid curves are the theoretical results of
Eqs. \eqref{velocity} and \eqref{decay}. Inset: Ratio $\protect\omega_c/
\protect\gamma h_0$ versus $\protect\alpha $. Solid curve is the theoretical
formulation $\frac{\protect\alpha }{1+\protect\alpha^2}$ as we get.}
\label{fig2}
\end{figure}

To test validity of our analytical results, we solve Eq. \eqref{LLG}
numerically for a uniaxial 1D wire of magnetic anisotropy $U=-\frac{1}{2}
K\cos^2\theta$. We first scale time, length, energy density and field
amplitude in units of $\left( \gamma M_{s}\right) ^{-1},$ $\sqrt{A/\mu
_0M_s^2},$ $\mu_0M_s^2$ and $M_s,$ respectively, so that velocity is in the
unit of $\gamma \sqrt{A/\mu_0}$. We then adopt a standard arithmetic
Method-of-Lines to discretize space with an adaptive time-step control. The
length of the wire is 100 and the mesh size is 0.2. The density plot of $m_z$
in $z-t$ plane determines the DW position. The DW velocity is extracted from
the slope of the DW position line. We find that the DW position line is a
straight line at low frequency, and it becomes an oscillation curve at high
frequency.

The symbols in Fig. 2 are numerical data for $K=8$, $\alpha =0.1$ and
various $h_{0}=3,\ 1$ and $0.5.$ It is clear that $h_{0}$ affects DW
velocity. The larger $h_{0}$ is, the higher the velocity will be, as given
by Eqs. \eqref{velocity} and \eqref{decay} because the DW width is increased
by a factor $1/(1-\sin \theta _{0})$. The agreement between Eq. %
\eqref{velocity} (straight lines in Fig. 2 \textit{without any fitting
parameters}) and numerical results for low-frequency are good. Eq. %
\eqref{decay} (curves in Fig. 2) captures qualitatively the decay feature
for large frequency, but systematically smaller than the numerical values.
This should not be a surprise since Eqs. \eqref{DWmotion} and %
\eqref{Oscillation} are derived from rigid DW assumption that, strictly
speaking, does not hold for $\omega >\omega _{c}$. To further test our
assumptions for Eqs. \eqref{DWmotion} and \eqref{Oscillation}, we compare
numerical values of $\omega _{c}/(\gamma h_{0})$ with theoretical prediction
of $\alpha /(1+\alpha ^{2})$ (curve in the inset of Fig. 2). The symbols
(squares, circles, and stars for $h_{0}=3,\ 1,$ and $0.5$, respectively) in
the inset of Fig. 2 are numerical data for various $\alpha $ that compares
well with the theoretical prediction.

It is predicted and confirmed\cite{Barnes,Shengyuan} that a DW motion with
its plane precessing around wire axis can generate an electromotive
potential of $\frac{\hbar}{e}\frac{d\phi}{dt}$ between the two sides of the
DW, where $\frac{d\phi }{dt}$ is the precession velocity of the DW. However,
a rigid body propagation along the wire will not generate an electromotive
force. A naive application of this theory to our DW motion seems lead to
zero electromotive force in the rotation frame and non-zero electromotive
force in the laboratory frame. This obvious contradiction is due to the
neglect of a Coriolis field of $\vec{B}=-\frac{\vec{\omega}}{\gamma }$ along
wire axis in the rotation frame. Thus, a conduction electron virtually moves
across the DW will experience an static potential gain of $\mathscr{E}_{s}=%
\frac{2\vec{\mu}\cdot \vec{\omega}}{\gamma }=\frac{\hbar \omega }{e}$,
exactly what is predicted by Niu's theory\cite{Shengyuan} in the laboratory
frame.

A realistic magnetic wire will not be completely symmetric around wire axis.
In order to show the robustness of the physics discussed here, we consider a
biaxial anisotropy field $\vec{h}_{an}=\frac{1}{\mu_0 M_s}\left( K_1m_z\hat{z%
}-K_2m_x\hat{x}\right),$ $K_1$ and $K_2$ describe the anisotropies along the
easy- and hard-axis, respectively. Generally speaking, the synchronization
would not be perfect in the presence of $K_{2}$ because local spin needs to
climb over an extra energy barrier when it follows the motion of a CPMF.
Fig. 3 shows the numerical results of velocity-frequency dependence of DW
with biaxial anisotropies, where the general features are very similar to
those in Fig. 2. It compares well with Eqs. \eqref{velocity} and %
\eqref{decay}. The magnitude of the peak velocity can be estimated by using
materials parameters of Co: $M_{s}=1.4\times 10^6A/m,$ $A=4\times
10^{-11}J/m,$ $K_1=5.2\times 10^5J/m^3,$ and $\alpha=0.1.$ We find the peak
velocity of $\upsilon =21.5m/s$ at the critical frequency $\omega_c=0.17GHz$
for $h_0=100Oe.$
\begin{figure}[tbph]
\begin{center}
\includegraphics[width=9cm, height=7cm]{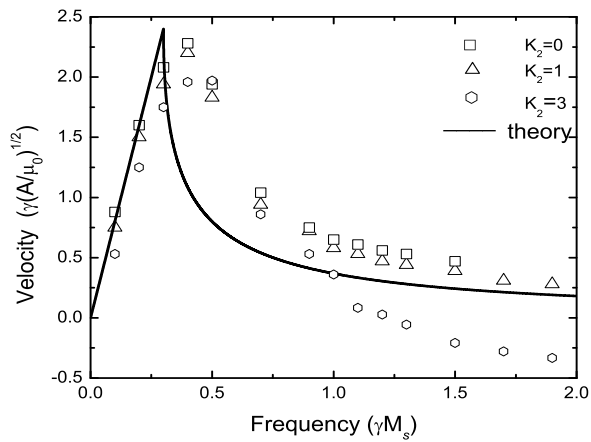}
\end{center}
\caption{DW propagation velocity $\protect\upsilon $ as a function of CPMF
frequency $\protect\omega$ for $\protect\alpha =0.1$, $h_0=3$ and $K_1=8$.
Symbols are results of numerical simulations for $K_2=0$ (squares), $1$
(triangles), and $3$ (hexagons), respectively. Solid curve corresponds to
the theoretical result for $K_2=0$.}
\label{fig3}
\end{figure}

In conclusion, CPMF at GHz frequency is an efficient control parameter for
DW motion. Two propagation modes are identified. Under a low frequency ($%
\omega <\omega _{c}$), a DW propagates like a rigid body at a constant
velocity that increases linearly with the CPMF frequency. At a high
frequency ($\omega >\omega _{c}$), DW propagation speed is oscillatory whose
time-averaged value decreases with the frequency. For a uniaxial wire, a DW
under a CPMF can be mapped to the DW under STT due to uniform spin-polarized
current. In the map, the CPMF frequency plays the role of the spin polarized
current.

This work is supported by Hong Kong UGC/CERG grants (\#603007, 603508 and
SBI07/08.SC09). We thank J. Lu for helping numerical simulations.


\begin{thebibliography}{99}
\bibitem{note1} Although various torques on individual spins are the
ultimate microscopic "forces" for DW dynamics. Energy may be a better
quantity to understand the origin of DW propagation. For detail, see
Reference 4.
\end{thebibliography}
\end{document}